\documentclass[fleqn,10pt]{wlscirep}
\usepackage[utf8]{inputenc}
\usepackage[T1]{fontenc}
\usepackage{textcomp}

\title{Individual skyrmion manipulation by local magnetic field gradients}

\author[1,2,*]{Arianna Casiraghi}
\author[2]{H\'ector Corte-Le\'on}
\author[3]{Mehran Vafaee}
\author[1]{Felipe Garcia-Sanchez}
\author[1]{Gianfranco Durin}
\author[1]{Massimo Pasquale}
\author[3]{Gerhard Jakob}
\author[3]{Mathias Kl\"aui}
\author[2]{Olga Kazakova}

\affil[1]{Istituto Nazionale di Ricerca Metrologica, Strada delle Cacce 91, 10135 Torino, Italy}
\affil[2]{National Physical Laboratory, Teddington TW11 0LW, United Kingdom}
\affil[3]{Institut f\"ur Physik, Johannes Gutenberg-Universit\"at Mainz, Staudinger Weg 7, 55128 Mainz, Germany}

\affil[*]{a.casiraghi@inrim.it}


\begin{abstract}

Magnetic skyrmions are topologically protected spin textures, stabilised in systems with strong Dzyaloshinskii-Moriya interaction (DMI). Several studies have shown that electrical currents can move skyrmions efficiently through spin-orbit torques. While promising for technological applications, current-driven skyrmion motion is intrinsically collective and accompanied by undesired heating effects. Here we demonstrate a new approach to control individual skyrmion positions precisely, which relies on the magnetic interaction between sample and a magnetic force microscopy (MFM) probe. We investigate perpendicularly magnetised X/CoFeB/MgO multilayers, where for X = W or Pt the DMI is sufficiently strong to allow for skyrmion nucleation in an applied field. We show that these skyrmions can be manipulated individually through the local field gradient generated by the scanning MFM probe with an unprecedented level of accuracy. Furthermore, we show that the probe stray field can assist skyrmion nucleation. Our proof-of-concepts results offer current-free paradigms to efficient individual skyrmion control.  
 
\end{abstract}
\begin{document}

\flushbottom
\maketitle

\thispagestyle{empty}

%\section*{Introduction}
Most ferromagnetic materials exhibit collinear magnetic ordering due to the Heisenberg exchange interaction between neighbouring spins. However, in magnetic systems with large spin-orbit coupling and lack of structural inversion symmetry, the Dzyaloshinskii-Moriya interaction (DMI) \cite{Dzyaloshinsky_JPhysChemSolids_1958, Moriya_PR_1960} can be strong enough to stabilise non-collinear spin textures, most notably magnetic skyrmions. Since their first discovery in chiral magnets \cite{Muhlbauer_Science_2009, Neubauer_PRL_2009}, magnetic skyrmions have attracted significant attention within the spintronics community due to their fascinating properties and potential applications \cite{Wiesendanger_NatRevMats_2016, Fert_NatRevMats_2017}. Indeed, their nanoscale dimensions combined with topological stabilisation \cite{Hagemeister_NatCommun_2015} and low threshold current densities required for motion \cite{Jonietz_Science_2010, Yu_NatCommun_2012, Fert_NatNanotechnol_2013, Iwasaki_NatNanotechnol_2013, Sampaio_NatNanotechnol_2013, Jiang_Science_2015}, make magnetic skyrmions promising candidates for storage and logic devices. In this technological context, particularly interesting are skyrmions found in systems composed of a heavy metal layer and an ultra-thin magnetic film with perpendicular magnetic anisotropy (PMA) \cite{Jiang_Science_2015, Moreau-Luchaire_NatNanotechnol_2016, Boulle_NatNanotechnol_2016, Woo_NatMater_2016, Legrand_NanoLett_2017, Litzius_NatPhysics_2017, Woo_NatCommun_2017}. Here the DMI arises due to the broken inversion symmetry at the interface and the large spin-orbit coupling in the heavy metal atoms, which mediate the interaction between spins in the ferromagnet. The interfacial nature of the DMI in these material stacks promotes the stabilisation of skyrmions of the N\'eel type \cite{Boulle_NatNanotechnol_2016}. Typical systems investigated utilise Co or CoFeB as magnetic films and Pt as heavy metal layer, while either an oxide or another heavy metal are used as top layers. In these material stacks, current pulses are usually employed to drive skyrmion motion, exploiting the spin Hall effect in the heavy metal layer and consequent spin-orbit torque \cite{Woo_NatMater_2016, Legrand_NanoLett_2017, Litzius_NatPhysics_2017, Woo_NatCommun_2017}. 
%skyrmions can be nucleated at room temperature by applying a perpendicular magnetic field in continuous films~\cite{} or a combination of magnetic field and current pulses in patterned structures~\cite{}. Current pulses are the most widely employed method to drive the motion of skyrmions, exploiting the spin Hall effect in the heavy metal layer and consequent spin-orbit torque~\cite{}. 
However, despite the low values of depinning currents required, fast skyrmion motion is only attainable using large current densities, which are accompanied by undesired Joule heating effects. Among the alternative methods proposed, using field gradients has been very recently shown to be an effective, current-free method to move skyrmion lattices in a chiral magnet \cite{Zhang_NatCommun_2018}. While a few theoretical studies have investigated this approach also for isolated skyrmions in thin-films \cite{Komineas_PRB_2015, Wang_NJP_2017, Liang_NJP_2018}, to the best of our knowledge no experimental work has yet been performed in these systems.

Here we show that isolated skyrmions can be manipulated individually and precisely using the local field gradient generated by a magnetic force microscopy (MFM) probe in ultra-thin films with PMA. The samples investigated consist of [X/CoFeB/MgO]$_{15}$ multilayers, with X = Ta, W or Pt. CoFeB/MgO was chosen due to its excellent homogeneity in terms of both thickness and PMA strength, which results in a low pinning energy landscape \cite{Burrowes_APL_2013, Woo_NatMater_2016}, making it a good candidate for reliable skyrmion motion. CoFeB is however a very soft magnetic material, meaning that imaging via MFM can be challenging due to potential perturbations caused by the probe stray field on the sample magnetic state (see Fig.~\ref{Fig1}). Indeed, magnetic stripe domains and their evolution into skyrmions, from which the DMI strength can also be extracted, have previously only been imaged at synchrotron facilities for these ultra-thin CoFeB layers, using non-interfering X-ray based methods \cite{Woo_NatMater_2016, Litzius_NatPhysics_2017, Woo_NatCommun_2017, Buttner_NatNanotechnol_2017, Jaiswal_APL_2017, Lemesh_AdvMater_2018}. In this work we show that a technique as straightforward and readily available as MFM can be used on CoFeB-based stacks at room temperature and in ambient conditions to: (i) determine the DMI strength through measurements of magnetic stripe-domain widths, (ii) image magnetic skyrmions and manipulate their motion with a high level of accuracy, and (iii) assist skyrmion nucleation in field.   
  
The labyrinth-like demagnetised state and its evolution with perpendicular field is imaged for all sample stacks, and the DMI constant $D$ is extracted from the average equilibrium domain width of the demagnetised state. %\textcolor{red}{, as well from the terminal domain width close to saturation}. 
Individual skyrmions are nucleated by the controllable shrinking of stripe domains in field. Since the probe-sample interaction is detrimental for imaging, we show that skyrmions can only be imaged in ``single pass'' MFM mode, with topography acquired prior to skyrmion nucleation. On the other hand, by exploiting the probe-sample interaction skyrmions can be moved in standard two pass MFM mode, through the local field gradient generated by the probe. This technique allows for the reversible and controllable manipulation of individual skyrmions, contrary to more conventional current-driven motion where skyrmions move in a collective manner. Finally, we show that the probe-sample interaction can also be used to create individual skyrmions at magnetic fields lower than their intrinsic nucleation field, by either collapsing stripe domains into skyrmions or by directly cutting through the stripe domains to generate skyrmions.   
% Since the probe-sample interaction is detrimental for imaging, the labyrinth-like demagnetised state and its evolution with field up to skyrmion nucleation is imaged taking several precautions. In particular, skyrmions are imaged in ``single pass'' MFM mode, with the topography acquired separately, prior to skyrmion nucleation. It is indeed by exploiting the strong probe-sample interaction that skyrmions can be moved in standard two pass MFM mode. We find that the local field gradient generated by the probe drives skyrmions along the slow scan axis direction. Skyrmion motion by this technique is reversible and highly controllable, also enabled by the low pinning landscape of CoFeB. Furthermore, skyrmion manipulation is attained at an individual level, contrary to more conventional current-driven motion where skyrmions move in a collective manner~\cite{}. 

\section*{Results and discussion}
\subsection*{Probe-sample interaction}
Since MFM was used to investigate the magnetic configuration for all sample stacks, as well as to image, nucleate and manipulate individual skyrmions, it is important to assess the effects of probe-sample interaction during MFM scanning. The principle behind standard MFM imaging relies on detecting the probe phase shift resulting from its interaction with the magnetic stray field generated by the sample. In particular, MFM contrast is directly proportional to the gradient of the force experienced by the probe during scanning. It follows that image perturbations caused by the probe-sample interaction are a common artefact in MFM imaging, particularly for relatively soft magnetic materials like CoFeB.
\begin{figure}[ht]
\centering
\includegraphics[width=\linewidth]{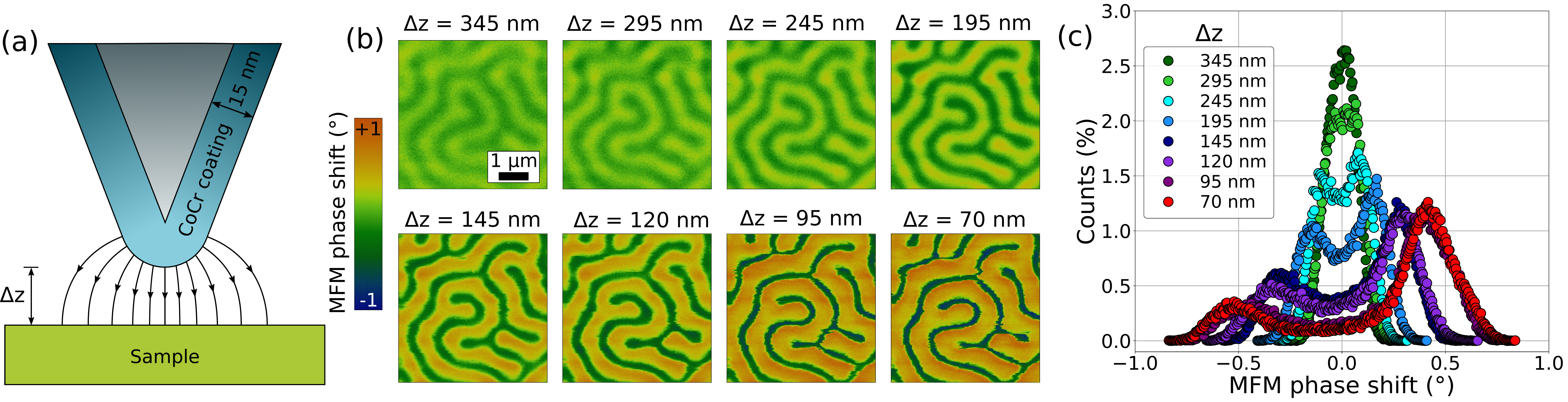}
\caption{\textbf{MFM image perturbations caused by probe-sample magnetic interaction.} (a) Schematic illustration of the magnetic stray field generated by an MFM probe. All MFM imaging was performed using low moment tips (NT-MDT MFM-LM), with magnetic CoCr coating thickness of $\sim$ 15 nm. $\Delta z$ represents the tip lift height used to record the MFM phase signal. (b) MFM imaging of the demagnetised state of Ta/CoFeB/MgO multilayer at different $\Delta z$, with the corresponding image histograms shown in (c). The bin width is kept constant across all histograms.}
\label{Fig1}
\end{figure}

Probe-induced modifications are represented clearly in Fig. \ref{Fig1}b,c, which show the demagnetised state for Ta/CoFeB/MgO imaged at different tip lift heights $\Delta z$ and the corresponding image histograms, respectively. While the contrast of domains magnetised up and down increases significantly upon reducing $\Delta z$, the influence of the tip, which has a stray field of $\sim$ 140 mT at its apex (see Supplementary Note 1 and Supplementary Figure 1), becomes more evident since the domains magnetised in the same direction as the local tip stray field expand at the expense of the others. This is reflected by an imbalance in the histogram peaks corresponding to the two domain types, which is visible already below $\Delta z$ as high as 245 nm. Significant domain distortions caused by the probe were also found in W/CoFeB/MgO, albeit of a reduced amount, while a much smaller effect was visible for the harder Pt/CoFeB/MgO sample even down to $\Delta z$ = 70 nm (see Supplementary Figures 2 and 3). This is consistent with magnetometry measurements which indicate that the perpendicular saturation field is lowest for Ta/CoFeB/MgO ($\sim$ 27 mT), intermediate for W/CoFeB/MgO ($\sim$ 36 mT) and highest for Pt/CoFeB/MgO ($\sim$ 94 mT) (see Supplementary Figure 4).

Distortions of the magnetic configuration caused by the probe-sample interaction are evidently detrimental for imaging, and precautions have to be taken in order to image the samples in the demagnetised state as well as to image individual skyrmions. Later we show that the same probe-sample interaction can actually be employed advantageously to control skyrmion motion.

\subsection*{DMI measurements}
The DMI constant $D$ is known to reduce the domain wall energy density of  N\'eel walls $\sigma_{DW}$, following the relationship\cite{Heide_PRB_2008, Thiaville_EPL_2012}: 
\begin{equation}
\sigma_{DW} = 4 \sqrt{AK_{eff}} - \pi \vert D \vert,
\label{eq1}
\end{equation}
where $A$ is the exchange stiffness and $K_{eff}$ is the effective perpendicular anisotropy constant. Accordingly, domain formation becomes more favourable in the presence of a finite DMI and measurements of the domain width can then be used to directly quantify $D$ \cite{Woo_NatMater_2016, Lemesh_PRB_2017}. In particular, following the approach implemented by Woo \textit{et al.} \cite{Woo_NatMater_2016}, we extracted $D$ from the equilibrium domain width of the demagnetised state, where up and down domains have the same average size.%\textcolor{red}{, as well as from the terminal domain width in a field close to saturation, when one of the two types of domains is about to disappear}. 

\begin{figure}[ht]
\centering
\includegraphics[width=\linewidth]{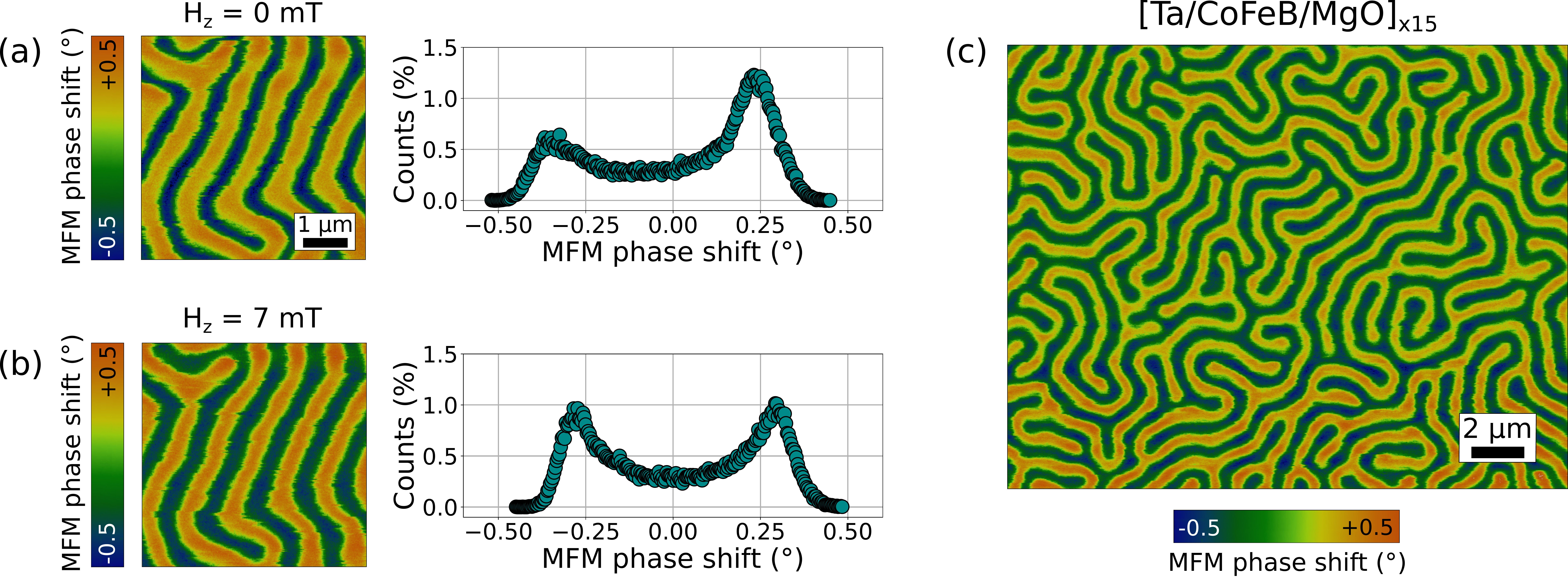}
\caption{\textbf{Procedure used to image the demagnetised configuration.} MFM imaging of Ta/CoFeB/MgO multilayer at $\Delta z$ = 145 nm without (a) and with (b) an external perpendicular bias field $H_z$ $\sim$ 7 mT, applied to compensate for the probe stray field. Image histograms are shown beside the corresponding MFM images. (c) Example of one of the MFM images for the Ta/CoFeB/MgO stack used to extract the DMI constant $D$. $\Delta z$ and $H_z$ are the same as in (b).}
\label{Fig2}
\end{figure}

Since the MFM probe alters significantly the relative domain population when scanning at $\Delta z$ low enough to grant a good magnetic contrast (see Fig. \ref{Fig1}), the demagnetised state of the samples was imaged while simultaneously applying an out-of-plane external bias field to counteract the perturbing effect of the tip. Only in this way it became possible to achieve an equal balance between up and down magnetised domains and image the demagnetised configuration as close as possible to an unperturbed state. This is schematically illustrated in Fig. \ref{Fig2}a,b for the Ta/CoFeB/MgO multilayer: by scanning at $\Delta z$ = 145 nm, an equal domain distribution is restored upon application of a perpendicular field $H_z$ $\sim$ 7 mT.       
In general, it was found that varying strengths of this bias field $H_z$ $\sim$ 7 $\pm$ 2 mT were needed to offset the local probe stray field at $\Delta z$ = 145 nm, which was attributed to variations among tips or their ageing stage. This value is in good agreement with the stray field value calculated through the tip transfer method, as explained in Supplementary Note 1. One of the MFM images for the demagnetised state of the Ta/CoFeB/MgO multilayer, that was analysed to quantify $D$, is shown in Fig. \ref{Fig2}c. Similar images for the W-based and Pt-based stacks are displayed in Supplementary Figure 5. The labyrinth domain structure observed in all samples is typical of multilayers with PMA and is also consistent with the slanted perpendicular hysteresis loops (see Supplementary Figure 4).  

Two different methods are used to extract the equilibrium domain width from the demagnetised images. The first approach relies on performing a 2-D fast Fourier transform (FFT) which, averaged across its radial profile, yields directly an amplitude peak at the reciprocal wavelength corresponding to the average domain periodicity \cite{Yamanouchi_IEEEMagnLett_2011}. The second approach uses an image segmentation process based on the random walker algorithm \cite{Grady_2006} to accurately separate pixels belonging to domains magnetised up or down. From this, the average domain width for each domain type is calculated as the ratio between the total area covered by one domain type and half of the total length of all domain edges detected (see Supplementary Note 2 and Supplementyary Figure 6). The average equilibrium domain widths extracted across several images via these two different methods are in excellent agreement. Using the approach implemented by Woo \textit{et al.} \cite{Woo_NatMater_2016}, the domain wall energy density $\sigma_{DW}$ can then be extracted from the average equilibrium domain width according to the analytical domain-spacing model applicable to multilayers \cite{Malek_CzechJPhys_1958}. Finally, once $\sigma_{DW}$ is known, $D$ is easily calculated according to Eq. \ref{eq1}. For the exchange stiffness $A$ a value of 10 pJ/m was assumed. %(calculations of $D$ with $A$ = 20 pJ/m are presented in Supplementary Table 1). 
The DMI magnitude extracted for the different samples is summarised in Table \ref{Tab1}, together with the average equilibrium domain width, and the relevant magnetic parameters, which were measured through vibrating sample magnetometry (VSM).
\begin{table}[ht]
{
\center
\renewcommand{\arraystretch}{1.5}
\begin{tabular}{c|c|c|c|c|c}
\hline 
\hline
\textbf{Multilayer} & \textbf{M$_{\mathbf{s}}$} (A/m)& \textbf{K$_{\mathbf{eff}}$} (J/m$^{3}$) & \textbf{Eq. width (nm)} & $\vert$\textbf{$\mathbf{{D}}$}$\vert$ (mJ/m$^{2}$) & $\vert$\textbf{$\mathbf{{D_{s}}}$}$\vert$ (pJ/m)\\
\hline 
\hline
[Ta/CoFeB/MgO]$_{\times 15}$ & 1.00 $\times$ 10$^6$ & 4.6 $\times$ 10$^5$ & 445 $\pm$ 7 &  0.08 $\pm$ 0.03 & 0.08 $\pm$ 0.03 \\
\hline
[W/CoFeB/MgO]$_{\times 15}$ & 1.17 $\times$ 10$^6$ & 4.1 $\times$ 10$^5$ & 365 $\pm$ 7 &  0.61 $\pm$ 0.03 & 0.37 $\pm$ 0.02 \\
\hline
[Pt/CoFeB/MgO]$_{\times 15}$ & 1.54 $\times$ 10$^6$ & 2.9 $\times$ 10$^5$ & 177 $\pm$ 4 &  1.0 $\pm$ 0.1 & 0.80 $\pm$ 0.08 \\
\hline 
\hline
\end{tabular}
\caption{\textbf{DMI values extracted from average equilibrium domain widths.} Saturation magnetisation $M_{s}$, effective perpendicular anisotropy constant $K_{eff}$, average equilibrium domain width (calculated as the average between the results of the two image analysis methods applied to several different images), absolute value of the DMI constant $D$ and of $D_{s} =  D \cdot t$, where $t$ is the CoFeB thickness, are listed for all the samples investigated. The values of $\vert{D}\vert$, and $\vert{D_s}\vert$ are calculated assuming $A$ = 10 pJ/m. The errors in $\vert{D}\vert$ and $\vert{D_s}\vert$ derive from the errors on estimating the average equilibrium domain width.} 
\label{Tab1}
}
\end{table}

The magnitude of $D_{s}$ extracted for Ta/CoFeB/MgO and W/CoFeB/MgO multilayers is in line with values reported for similar systems, in experiments that quantify the DMI through current-driven or field-driven domain wall motion, as well as from measurements of equilibrium domain patterns \cite{Torrejon_NatCommun_2014, LoConte_PRB_2015, Khan_APL_2016, Jaiswal_APL_2017}. On the other hand, lower estimations of $D_{s}$ in these systems have been extracted through Brillouin Light Scattering (BLS) measurements \cite{Soucaille_PRB_2016, Ma_PRL_2018}. Regarding the Pt/CoFeB/MgO multilayer our measured $D_{s}$ is lower than what observed in similar multilayers through measurements of equilibrium domain patterns \cite{Litzius_NatPhysics_2017, Buttner_NatNanotechnol_2017, Lemesh_AdvMater_2018}, but in good agreement with BLS experiments performed on individual layers \cite{Di_APL_2015, Ma_PRL_2018, Chen_APL_2018}. In any case, only for W/CoFeB/MgO and Pt/CoFeB/MgO multilayers the DMI magnitude is significantly larger than the critical amount required to stabilise N\'eel domain walls \cite{Thiaville_EPL_2012}, with $\vert D_{c} \vert = 2 \mathrm{ln}(2) \mu_{0} M_{S}^2t / \pi^2$ being 0.18 mJ/m$^{2}$, 0.15 mJ/m$^{2}$, and 0.33 mJ/m$^{2}$ for Ta, W and Pt-based stacks, respectively. Furthermore, the DMI magnitude is always lower than the threshold value for spontaneous skyrmion generation \cite{Rohart_PRB_2013}, $\vert D_{th} \vert = 4 \sqrt{A K_{eff}} / \pi$, indicating that skymrion lattices or individual skyrmions cannot be generated in these systems without some form of external excitation. Indeed, as will be shown in the next section, skyrmions could only be stabilised in W/CoFeB/MgO and Pt/CoFeB/MgO multilayers by applying large perpendicular magnetic fields.

\subsection*{Skyrmion imaging and manipulation}
For all investigated samples, application of a perpendicular magnetic field results in the contraction of stripe domains that are magnetised against the field direction. In systems with a large enough DMI these shrinking stripe domains can collapse into stable individual skyrmions upon increasing the perpendicular field towards saturation \cite{Moreau-Luchaire_NatNanotechnol_2016, Woo_NatMater_2016, Woo_NatCommun_2017, Soumyanarayanan_NatMater_2017}. This was observed in W/CoFeB/MgO and Pt/CoFeB/MgO multilayers, while the small DMI in the Ta-based stack prevented skyrmions from being stabilised in field. 

\begin{figure}[ht]
\centering
\includegraphics[width=\linewidth]{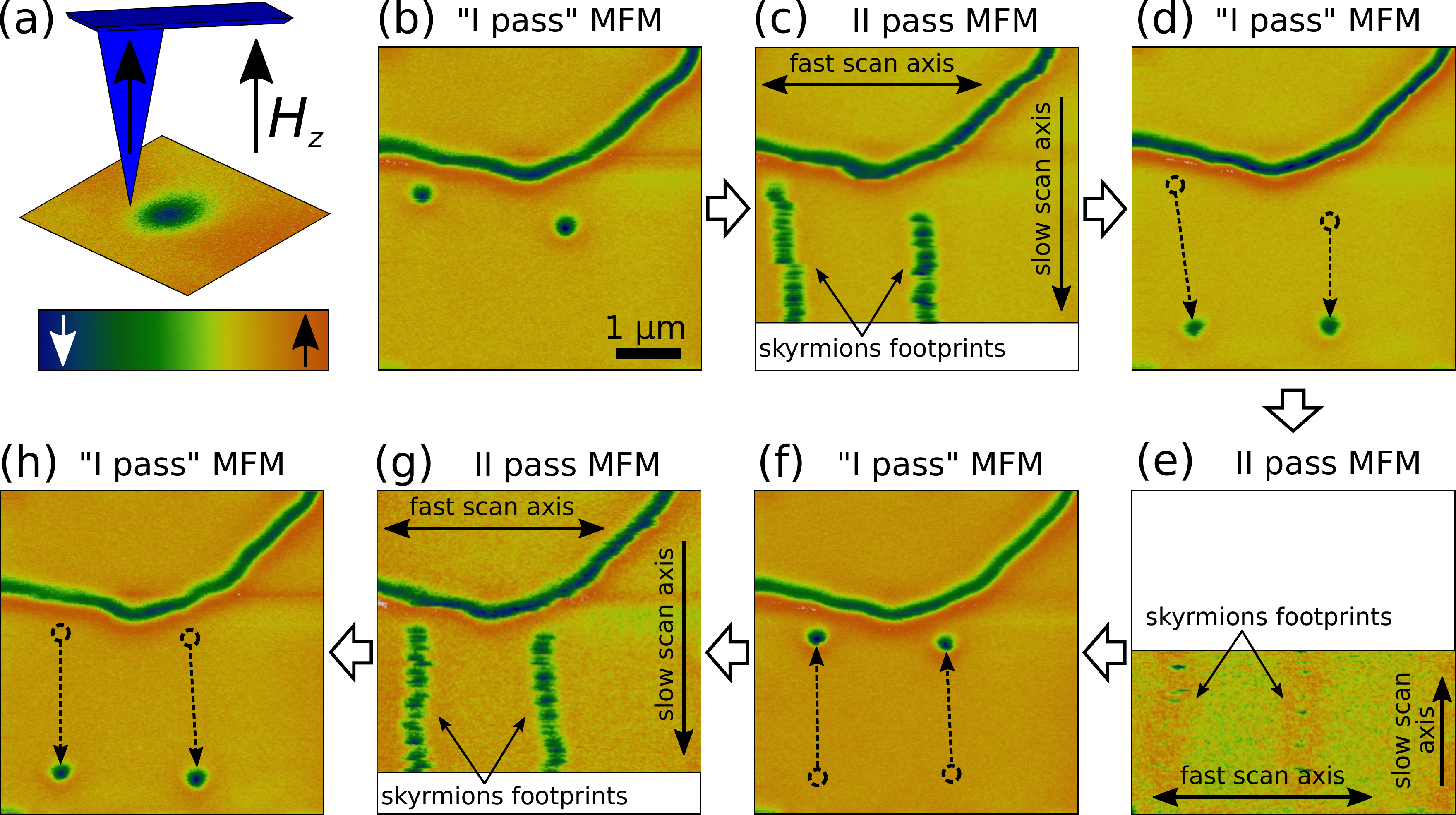}
\caption{\textbf{MFM imaging of individual skyrmions and manipulation by local field gradients.} Sequence of MFM images illustrating individual skyrmion manipulation by MFM in Pt/CoFeB/MgO multilayer under an applied perpendicular field $H_z$ = 71 mT. (b,d,f,h) Skyrmions are imaged in single pass MFM mode at a lift height $\Delta z$ = 145 nm, with the topography acquired separately at $H_z$ = 50 mT. The dashed arrows are guides for the eye indicating skyrmion motion between consecutive steps. (c,e,g) Skyrmions are moved by the probe stray field during the topography line scans of a standard two pass MFM scan (also with $\Delta z$ = 145 nm), which reveal the trajectory followed by skyrmions in the form of actual skyrmion ``footprints''. These scans were stopped before reaching the end only to prevent skyrmions from moving outside of the selected region. The directions of fast (with trace and retrace line scans) and slow MFM scan axes are marked in the individual images. The spacing between consecutive line scans along the slow scan axis is $\delta$ $\sim$ 20 nm. As indicated by the sketch in (a), the tip is magnetised along the external field direction for all MFM images, thus opposing the skyrmions core magnetisation.}
\label{Fig3}
\end{figure}

In order to image the nucleation of skyrmions, standard two pass MFM imaging was initially attempted. However, the magnetic interaction between probe and sample during first pass line scans, when sample topography is recorded, was found to be too strong to allow for skyrmion imaging using this MFM mode, even when using commercial low moment field probes. As a matter of fact, individual skyrmions could only be imaged in ``single pass'' MFM mode, having acquired the topography of the whole region of interest separately (see Methods). In particular, the topography was recorded at a field below skyrmion nucleation, where stripe domains are still present and able to better withstand strong perturbations caused by probe-sample interactions. Upon increasing the field up to skyrmion nucleation, it becomes then possible to image unperturbed individual skyrmions by using the MFM in single pass mode, to record the phase signal only, at a set lift height. This is shown for Pt/CoFeB/MgO in Fig. \ref{Fig3}b, where two skyrmions and a long stripe domain are imaged in single pass MFM at 71 mT, after having recorded the topography of the corresponding area at 50 mT. The probe is magnetised in the same direction as the external field, as shown by the sketch in Fig. \ref{Fig3}a. 

Skyrmion nucleation occurred at 32 mT and 57 mT for W/CoFeB/MgO and Pt/CoFeB/MgO multilayers, respectively. The skyrmion nucleation field was found to be highly reproducible, also across different sample regions. Furthermore, the position of the individual skyrmions was observed to vary from one nucleation cycle to another, indicating that pinning does not play a significant role in the skyrmion nucleation process. This is likely to be enabled by the energy landscape typical of CoFeB which features lower pinning than Co, as shown for both domain wall motion \cite{Burrowes_APL_2013} and skyrmion motion \cite{Woo_NatMater_2016}. 
     
\begin{figure}[ht]
\centering
\includegraphics[width=0.5\linewidth]{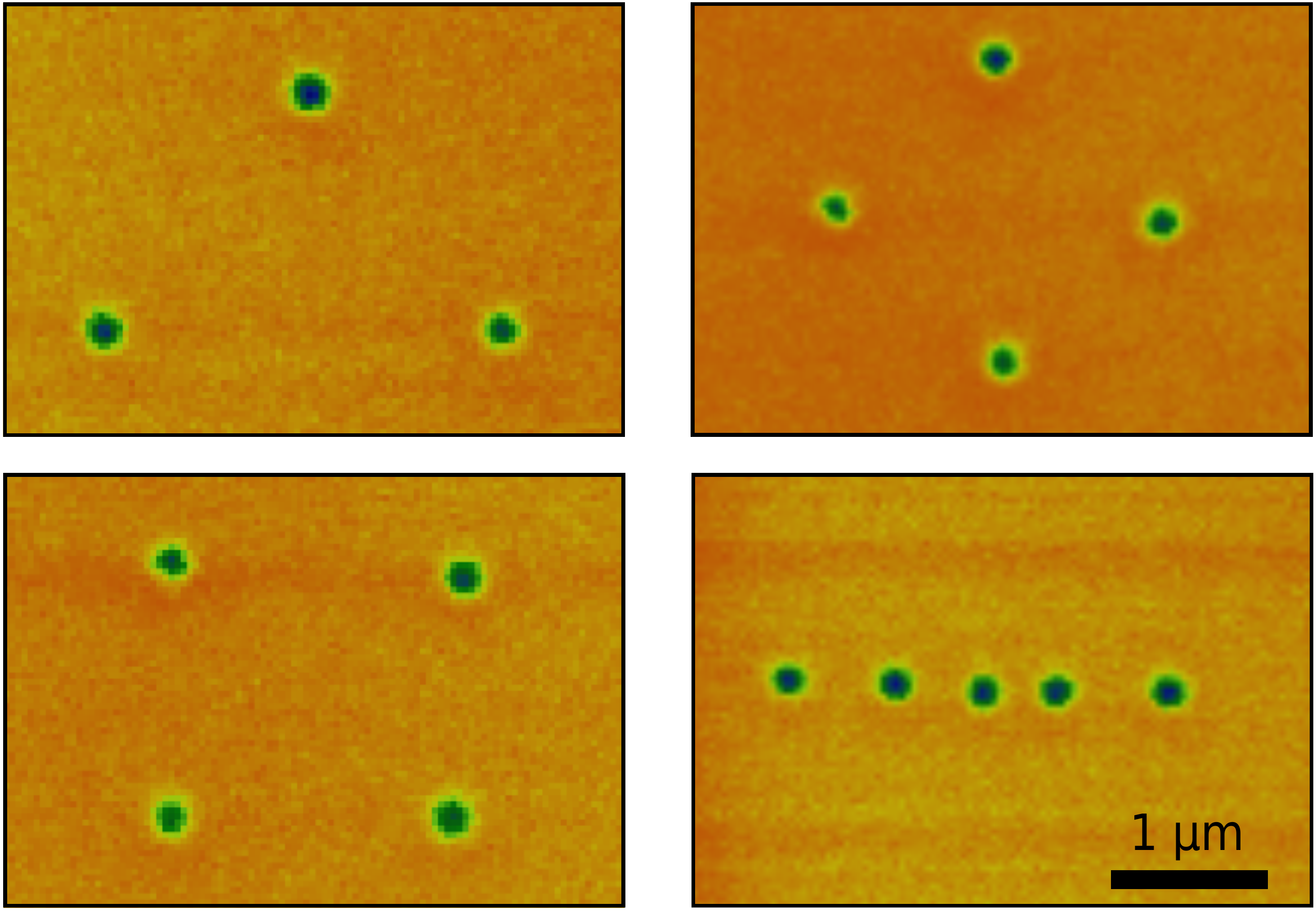}
\caption{\textbf{Drawing arbitrary patterns with individual skyrmions.} Single pass MFM imaging of different geometrical skyrmion patterns in Pt/CoFeB/MgO multilayer obtained by manipulating the individual skyrmions through a series of two pass MFM scans (not shown) with $\delta$ $\sim$ 10-20 nm. All images are taken at a lift height $\Delta z$ = 145 nm and under an applied perpendicular field $H_z$ = 71 mT, with the topography acquired separately at $H_z$ = 50 mT. The tip is magnetised along the external field direction, thus opposing the skyrmions core magnetisation.}
\label{Fig4}
\end{figure}

Although detrimental for imaging skyrmions, the probe-sample interaction can actually be exploited to move skyrmions efficiently. This is illustrated for Pt/CoFeB/MgO in Fig. \ref{Fig3}c, which was acquired in standard two pass MFM, with the probe magnetised along the external field direction: during each line scan in first pass, because of the close proximity between probe and sample, the field gradient generated by the tip causes skyrmions to move further down, while during the second pass what is perceived as magnetic contrast is effectively the skyrmion trajectory. A comparison between Fig. \ref{Fig3}b and Fig. \ref{Fig3}d, which image the skyrmions before and after the two pass MFM scan, respectively, indicates that skyrmions have moved downwards by a few micrometers. Several repetitions of the single pass scan in Fig. \ref{Fig3}d were acquired to ensure the stability of the new skyrmion position. It is interesting to observe that only skyrmions are moved by the MFM probe, while the position of the stripe domain visible at the top of the images in Fig. \ref{Fig3} is more stable against the local tip stray field. Motion of skyrmions occurs along the direction of the MFM slow scan axis, i.e. skyrmions move downwards when scanning in two pass mode from top to bottom. Therefore, skyrmion motion can easily be reversed by inverting the slow scan axis direction, as shown in the sequence of images in Fig. \ref{Fig3}. 
%On the other hand, skyrmion motion is unaffected by the direction of the fast scan axis: whether the tip is allowed to perform both trace and retrace line scans (i.e. left to right and viceversa, as in Fig. \ref{Fig3}) or only one of them, skyrmions are always driven roughly orthogonally to the the fast scan axis. 
Furthermore, it is important to note that the spacing $\delta$ between consecutive line scans along the slow scan axis should be kept small in order to attain reliable skyrmion motion. In our case, skyrmions could be moved effectively for $\delta$ $\sim$ 10-20 nm, while increasing the spacing to $\delta$ $\sim$ 40 nm or above resulted in stationary skyrmions or skyrmions that could only be moved a little before decoupling from the scanning tip. Arguably, a reliable motion cannot be expected if $\delta$ becomes comparable to the size of the skyrmion core. By using $\delta$ $\sim$ 10-20 nm in two pass MFM scans, individual skyrmion motion can be controlled with a high level of accuracy and the MFM tip can effectively become a tool to manipulate skyrmions in any desired manner. This is demonstrated in Fig. \ref{Fig4}, where arbitrary geometrical patterns have been drawn with skyrmions, by moving them from random positions through a series of two pass MFM scans with different slow scan axis directions. 

\begin{figure}[ht]
\centering
\includegraphics[width=0.75\linewidth]{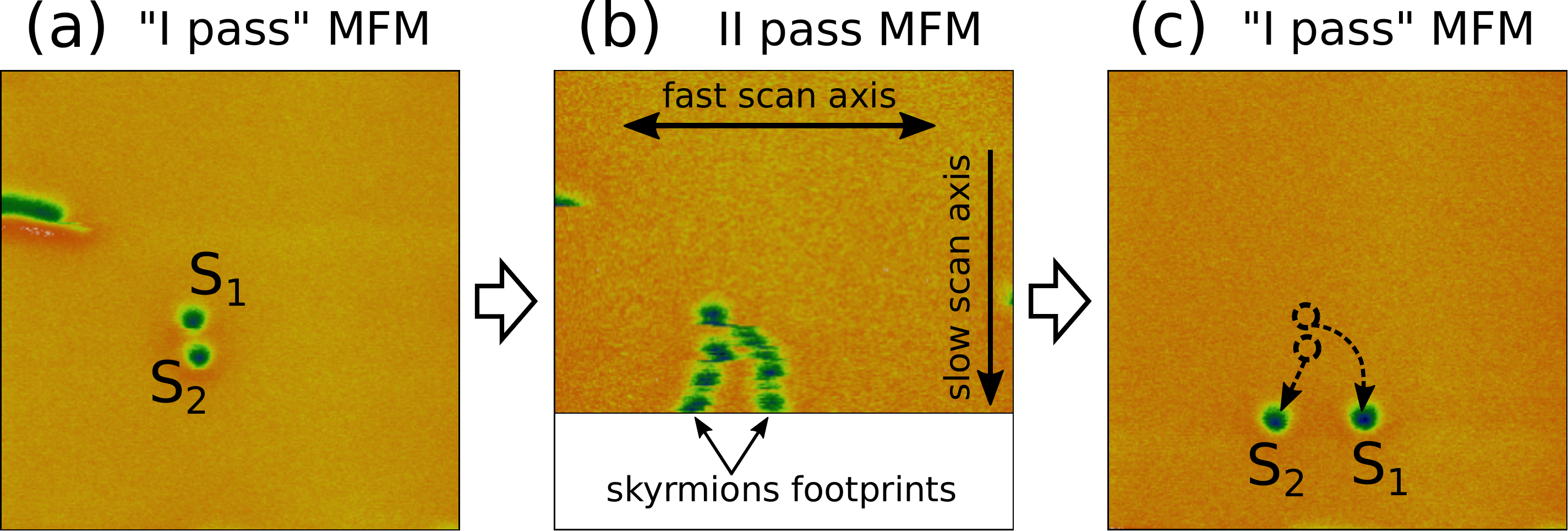}
\caption{\textbf{MFM imaging of the probe-driven interaction between two individual skyrmions.} Sequence of MFM images illustrating the dynamics of probe-driven skyrmion interaction in Pt/CoFeB/MgO multilayer under an applied perpendicular field $H_z$ = 71 mT. (a,c) Two individual skyrmions, marked by S$_1$ and S$_2$, are imaged in single pass MFM mode at a lift height $\Delta z$ = 145 nm, with the topography acquired separately at $H_z$ = 50 mT. The dashed arrows are guides for the eye indicating the trajectories followed by the skyrmions during the probe-driven interaction. (b) Skyrmion S$_1$ is pushed against skyrmion S$_2$ by the probe stray field during the topography line scans of a standard two pass MFM scan (also with $\Delta z$ = 145 nm). Skyrmion S$_1$ avoids the collision with skyrmion S$_2$ by jumping abruptly towards the right. The directions of fast and slow scan axes are marked by the arrows. The tip is magnetised along the external field direction, thus opposing the skyrmions core magnetisation.      
}
\label{Fig5}
\end{figure}

Finally, we investigate the dynamics of two adjacent skyrmions when being pushed against each other by the MFM probe. As illustrated in Fig. \ref{Fig5}a, two individual skyrmions are initially positioned by the MFM tip next to each other, with skyrmion S$_1$ being directly above skyrmion S$_2$. When performing a two pass MFM scan from top to bottom, the local field gradient generated by the probe drives S$_1$ towards S$_2$. The two skyrmions ``footprints'' are visible in Fig. \ref{Fig5}b and indicate that S$_1$, when being pushed downwards, avoids interacting with S$_2$ by suddenly jumping towards the right. Analogously, S$_2$ moves slightly towards the left. The final skyrmions positions are illustrated in Fig. \ref{Fig5}c.        
The observed repulsive interaction between the two skyrmions is an indication of their topological nature and of the topological barrier stabilising their structure, which differentiate skyrmions from topologically trivial magnetic bubbles \cite{Nagaosa_NatNanotechnol_2013, Hagemeister_NatCommun_2015, Wiesendanger_NatRevMats_2016, Fert_NatRevMats_2017}. %\textcolor{red}{This circumstance is also reinforced by the size}

\subsection*{Skyrmion nucleation at lower magnetic fields}
\begin{figure}[ht]
\centering
\includegraphics[width=\linewidth]{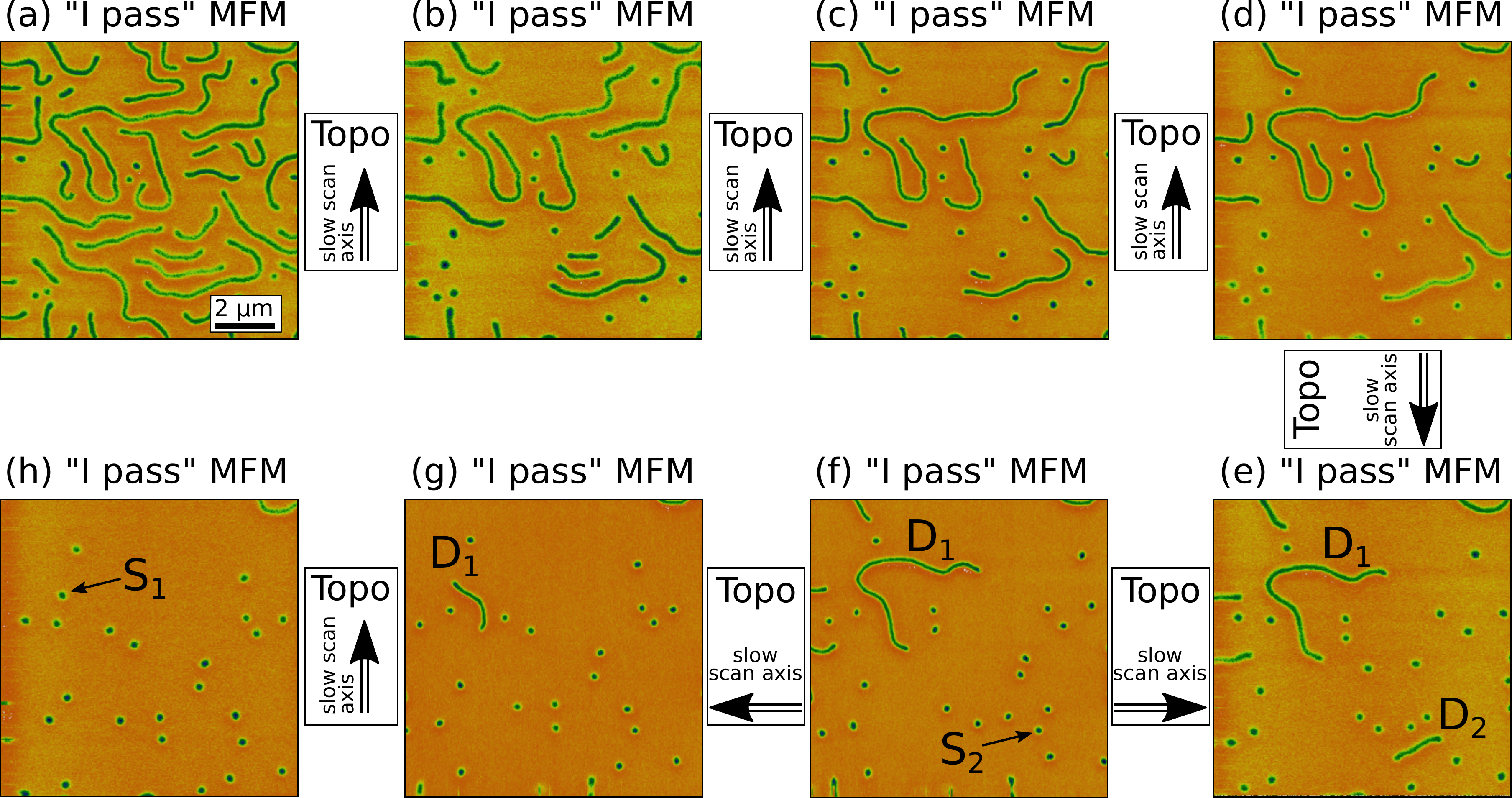}
\caption{\textbf{MFM probe-induced collapse of domains into individual skyrmions.} Sequence of MFM images illustrating conversion of domains into skyrmions via MFM in Pt/CoFeB/MgO multilayer under an applied perpendicular field $H_z$ = 57 mT. All images are acquired in single pass MFM mode at a lift height $\Delta z$ = 145 nm. A topography scan with a line spacing $\delta$ $\sim$ 40 nm is performed between each image (not shown), with the slow scan axis direction as indicated by the arrows. Symbols D$_{1}$ and D$_{2}$ mark two stripe domains which convert into skyrmions S$_{1}$ and S$_{2}$, respectively, upon successive topography scans, as explained in the text. In all images the tip is magnetised along the external field direction, thus opposing the skyrmions core magnetisation.}
\label{Fig6}
\end{figure}

We already mentioned that the skyrmion nucleation field for Pt/CoFeB/MgO was reproducibly found to be 57 mT. At this field, a few individual skyrmions are stabilised, while many long stripe domains are still present in a typical 10 $\times$ 10 $\mu$m$^{2}$ image (see Fig. \ref{Fig6}a). In order to convert all stripe domains into skyrmions, fields higher than 75 mT need to be applied. Here we show that the probe stray field can be used to promote skyrmion nucleation by assisting domain contraction without increasing the external field. This is illustrated through the sequence of MFM single pass images shown in Fig. \ref{Fig6}. 
Between each image a topography scan is performed (not shown) with the direction of the slow scan axis as indicated. Many elongated domains are contracted into skyrmions after only one topography scan (see Fig. \ref{Fig6}b), but repeating the topography scan in the same direction does not change significantly further the magnetic configuration, as several long domains remain unvaried (see Figs. \ref{Fig6}c,d). However, it is found here that even these more persistent domains can collapse into skyrmions by changing the slow scan axis direction in such a way that the scanning tip moves against one of the domain extremities. For instance, the topography between Figs. \ref{Fig6}e and \ref{Fig6}f is scanned from left to right and leads to the contraction of stripe domain D$_{2}$ into skyrmion S$_{2}$, while leaving stripe domain D$_{1}$ almost unaffected. To convert D$_{1}$ into a skyrmion, the topography has to be scanned firstly from right to left, with the tip moving against the extremities of D$_{1}$, leading to a shorter vertical stripe domain (see Fig. \ref{Fig6}g), and then from bottom to top, finally resulting in skyrmion S$_1$ (see Fig. \ref{Fig6}h). Thus, this method allows to convert all stripe domains into skyrmions without increasing the external field, and more efficiently than what would otherwise be possible by keeping the same direction of the slow scan axis. It is important to note that skyrmions in Fig. \ref{Fig6} do not move during topography scans in the same reliable and controllable manner that was shown in Fig. \ref{Fig3}. This is due to the fact that a larger line spacing $\delta$ $\sim$ 40 nm was used in Fig. \ref{Fig6}, resulting in small or negligible skyrmion motion, as already explained.

\begin{figure}[ht]
\centering
\includegraphics[width=0.6\linewidth]{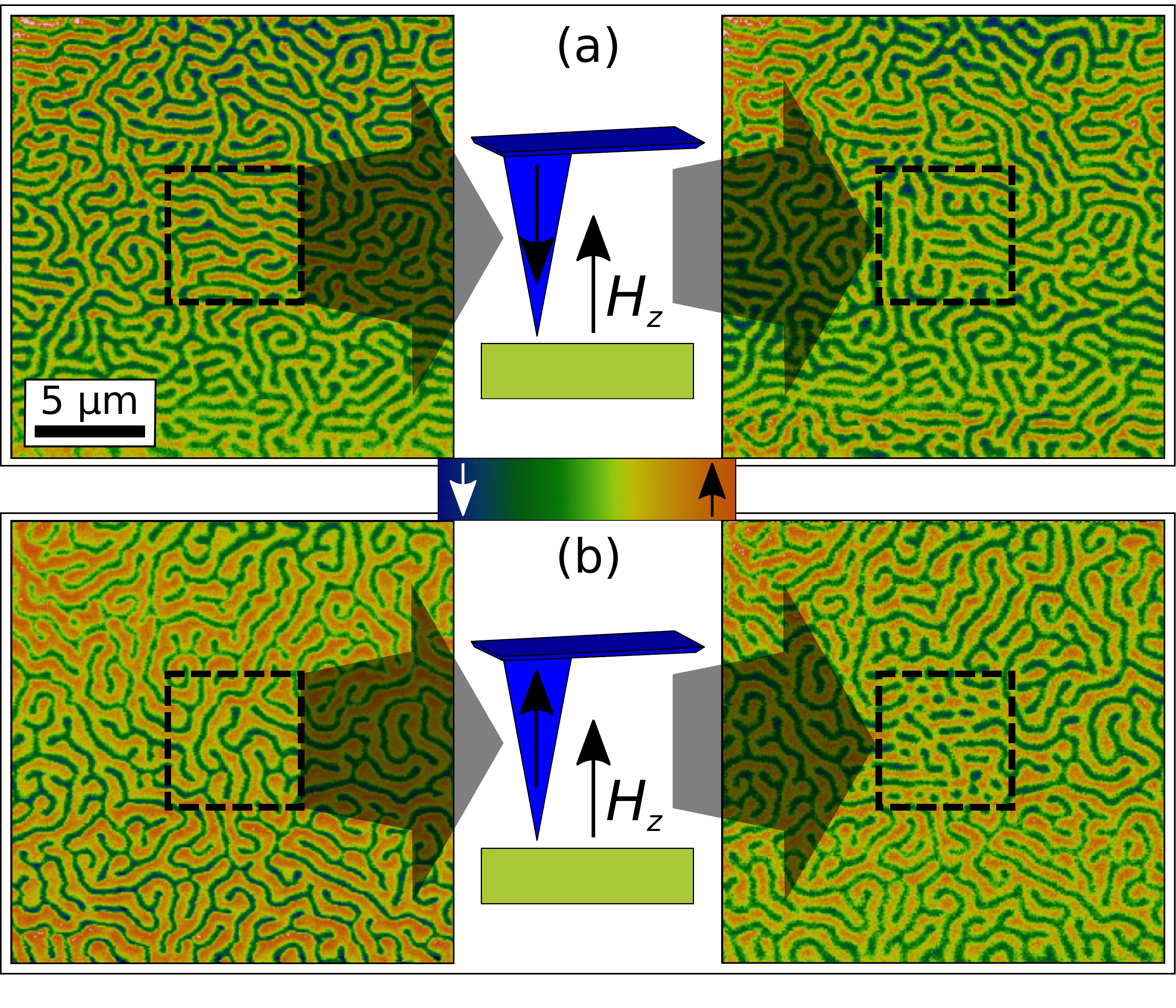}
\caption{\textbf{MFM probe-cutting of labyrinth domains into individual skyrmions.} Single pass MFM images of W/CoFeB/MgO multilayer under an applied field of $\sim$ 10 mT before (left) and after (right) a topography scan of the central area (not show) with the tip magnetised against (a) or along (b) the external field. The topography scan was performed with the slow scan axis from bottom to top. The color scale bar indicates the direction of the magnetisation in the sample.} 
\label{Fig7}
\end{figure}

Finally, we discuss the possibility of nucleating skyrmions by ``cutting'' the stripe domains with the tip stray field. To this end, it is important to note that the switching field of the tip magnetisation was measured to be $\sim$ 40-50 mT (see Supplementary Figure 7), which means that when applying external fields larger than this value, the tip is always magnetised along the field direction. This is the case for all the MFM images presented for Pt/CoFeB/MgO thus far (i.e. Figs. \ref{Fig3}-\ref{Fig6}), whereby external field and tip stray field are aligned in the same direction, opposing the magnetisation of the skyrmions core. Skyrmion nucleation through domain slicing was first attempted in these conditions for Pt/CoFeB/MgO, by scanning the tip several times in topography mode across a few stripe domains, under an applied perpendicular field $H_z$ = 57 mT. Despite external field and tip stray field being aligned with each other, no visible modification to the sample magnetic configuration was obtained.   
%To this end, it is important to note that the switching field of the tip magnetisation was measured to be $\sim$ 45-50 mT (see Supplementary Info), therefore intermediate between the skyrmion nucleation field for W/CoFeB/MgO and Pt/CoFeB/MgO. This means that for Pt/CoFeB/MgO the tip is always magnetised in the same direction as the applied field (and against the skyrmion core magnetisation) when imaging or manipulating skyrmions, while for W/CoFeB/MgO the influence of two tip configurations can be investigated, namely tip magnetised along or against the external field. 
However, for the softer W/CoFeB/MgO only one topography scan in a modest external field ($\sim$ 10 mT) was sufficient to significantly alter the labyrinth-like magnetic state. This is shown in Fig. \ref{Fig7} for two possible magnetic configurations of the tip. 
With the tip magnetisation aligned against the external field, the topography scan mainly leads to a rearrangement of the labyrinth domains (see Fig. \ref{Fig7}a), with no significant skyrmion nucleation. On the other hand, when the tip magnetisation is in the same direction as the field, so that the tip stray field adds to the external field, some of the dark stripe domains are effectively sliced through by the scanning tip and individual skyrmions are created (see Fig. \ref{Fig7}b), which is also consistent with previous work \cite{Zhang_APL_2018}. %\textcolor{red}{compare the tip stray field at the apex with saturation field of W/CoFeB/MgO as they do in ZhangAPL2018.}

\section*{Conclusions}
While MFM is well established as a tool to measure DMI and investigate nanoscale magnetic skyrmions in Co-based multilayer stacks \cite{Legrand_NanoLett_2017, Soumyanarayanan_NatMater_2017, Bacani_2018,  Maccariello_NatNanotechnol_2018, Zhang_APL_2018}, similar studies for CoFeB-based systems have involved using less widely available techniques, such as Scanning Transmission X-ray Microscopy (STXM) \cite{Woo_NatMater_2016, Litzius_NatPhysics_2017, Jaiswal_APL_2017, Lemesh_AdvMater_2018}, Magnetic Transmission soft X-ray Microscopy (MTXM) \cite{Woo_NatMater_2016, Woo_NatCommun_2017} or X-ray holography \cite{Buttner_NatNanotechnol_2017}. MFM imaging of soft magnetic materials like CoFeB can actually be very challenging due to image deformations caused by the probe-sample magnetic interaction. However, in this work we performed extensive MFM imaging of CoFeB/MgO multilayers, and we demonstrated that MFM can be used as a versatile technique to ascertain DMI and study magnetic skyrmions also in these systems. 

We determined the DMI from measurements of equilibrium domain widths in the demagnetised state. %\textcolor{red}{as well as of domain widths close to saturation}. 
To minimise the effect of tip-induced perturbations, commercial low moment tips were used, in combination with large lift heights. A small bias field was applied during imaging of the demagnetised state to compensate for the imbalance between up and down magnetised domains due to the probe stray field. Image analysis was then performed to extract the average equilibrium domain width with two different approaches, which yielded results in excellent agreement, and the DMI was then quantified through analytical domain-spacing model for multilayers.    

Furthermore, we explored how the probe-sample interaction can be beneficially exploited to attain individual skyrmion manipulation, as well as to assist skyrmion nucleation in field. In both cases, the local field gradient generated by the scanning MFM tip represents a tool not only to image skyrmions, but also to actively interact with them and manipulate their position controllably. Interestingly, while skyrmion imaging is only possible in ``single pass'' MFM scans, with the topography of the corresponding area acquired separately, standard two pass MFM provides a means to drive skyrmions (during topography scan, in first pass) and image their trajectory (during lift scan, in second pass). Skyrmion motion by this technique is reversible and highly controllable, also enabled by the low pinning landscape of CoFeB. Since the probe stray field opposes the direction of the skyrmions core magnetisation, we found that skyrmions move away from the scanning probe, in a direction approximately perpendicular to the probe fast scan axis (i.e. the trace/retrace axis). Furthermore, we show that the tip stray field in topography scans can also be used to promote the creation of skyrmions at fields lower than their intrinsic nucleation field by either collapsing labyrinth-like domains or directly cutting through them. Overall, our proof-of-concept results open alternative ways for the creation and manipulation of individual skyrmions in multilayers with perpendicular magnetic anisotropy.

\section*{Methods}
\subsection*{Sample growth}
The multilayer samples investigated consist of (all thicknesses in nm): (i) Si/SiO$_{2}$/[Ta(5)/Co$_{20}$Fe$_{60}$B$_{20}$(1)/MgO(2)]$_{15}$/Ta(5), (ii) Si/SiO$_{2}$/[W(5)/Co$_{20}$Fe$_{60}$B$_{20}$(0.6)/MgO(2)]$_{15}$/Ta(5), and (iii) Si/SiO$_{2}$/Ta(5.7)/[Pt(3.4)/Co$_{60}$Fe$_{20}$B$_{20}$(0.8)/MgO(1.4)]$_{15}$/Ta(5). They were grown at room temperature by d.c. magnetron sputtering in a sputter system with a base pressure of 5 $\times$ 10$^{-8}$mbar. After growth, the stacks were annealed in vacuum for 2 hours at a temperature of 300 $^{\circ}$C, for sample (i), and 400 $^{\circ}$C for samples (ii) and (iii). The 5 nm Ta capping layer was used to prevent sample oxidation. For sample (iii), the 5.7 nm thick Ta seed layer was used to improve adhesion between film and substrate.

\subsection*{Magnetometry}
The magnetic properties of the multilayers were studied using vibrating sample magnetometry (VSM) at room temperature. Both in-plane and out-of-plane magnetic hysteresis loops were measured (see Supplementary Figure 4). All investigated samples exhibit PMA, with a saturation field that increases in going from sample (i) to sample (iii). The saturation magnetisation $M_s$ and the effective perpendicular anisotropy constant $K_{eff}$, which was extracted from the in-plane saturation field, are listed in Tab. \ref{Tab1}. 

\subsection*{Magnetic force microscopy imaging}
MFM imaging of the multilayers was performed at room temperature with a NT-MDT Ntegra Aura scanning probe microscope (SPM). The system is fitted with an electromagnet which allows to apply out-of-plane magnetic fields up to $\sim$ 115 mT during scanning. Low moment tips (NT-MDT MFM-LM) were chosen in order to minimise probe-sample interaction. MFM imaging of the demagnetised state of the samples %\textcolor{red}{and of the terminal state} 
was carried out in standard two pass tapping mode, whereby for \textit{each scanned line} topography is acquired during the first pass, while the MFM phase signal of that line is acquired during the second pass at a set lift height $\Delta z$. 
%A small perpendicular field was applied during imaging to compensate for the distortions caused by the tip stray field.  
MFM imaging of individual skyrmions was instead performed in ``single pass'' tapping mode, which allows to acquire topography for the \textit{whole scanned area} first, followed by the MFM phase signal for the same area at a set lift height $\Delta z$. By working in single pass mode, it is then possible to measure topography and phase separately, which was essential to image skyrmions. Furthermore, single pass mode allows to acquire the phase signal at different magnetic fields without the need to lower the probe close to the sample to retake topography. Finally, individual skyrmion manipulation was attained using standard two pass tapping mode: skyrmions move due to the field gradient generated by the probe during first pass, when topography is recorded, while their trajectory is imaged during second pass.

\bibliography{biblio.bib}

\begin{thebibliography}{10}
\urlstyle{rm}
\expandafter\ifx\csname url\endcsname\relax
  \def\url#1{\texttt{#1}}\fi
\expandafter\ifx\csname urlprefix\endcsname\relax\def\urlprefix{URL }\fi
\expandafter\ifx\csname doiprefix\endcsname\relax\def\doiprefix{DOI: }\fi
\providecommand{\bibinfo}[2]{#2}
\providecommand{\eprint}[2][]{\url{#2}}

\bibitem{Dzyaloshinsky_JPhysChemSolids_1958}
\bibinfo{author}{Dzyaloshinskii, I.}
\newblock \bibinfo{journal}{\bibinfo{title}{Thermodynamic theory of "weak"
  ferromagnetism in antiferromagnetic substances}}.
\newblock {\emph{\JournalTitle{J. Phys. Chem. Solids}}}
  \textbf{\bibinfo{volume}{4}}, \bibinfo{pages}{241} (\bibinfo{year}{1958}).

\bibitem{Moriya_PR_1960}
\bibinfo{author}{Moriya, T.}
\newblock \bibinfo{journal}{\bibinfo{title}{Anisotropic superexchange
  interaction and weak ferromagnetism}}.
\newblock {\emph{\JournalTitle{Phys. Rev.}}} \textbf{\bibinfo{volume}{120}},
  \bibinfo{pages}{91} (\bibinfo{year}{1960}).

\bibitem{Muhlbauer_Science_2009}
\bibinfo{author}{M\"{u}hlbauer, S.} \emph{et~al.}
\newblock \bibinfo{journal}{\bibinfo{title}{Skyrmion lattice in a chiral
  magnet}}.
\newblock {\emph{\JournalTitle{Science}}} \textbf{\bibinfo{volume}{323}},
  \bibinfo{pages}{915} (\bibinfo{year}{2009}).

\bibitem{Neubauer_PRL_2009}
\bibinfo{author}{Neubauer, A.} \emph{et~al.}
\newblock \bibinfo{journal}{\bibinfo{title}{Topological {Hall} effect in the
  ${A}$ phase of {M}n{S}i}}.
\newblock {\emph{\JournalTitle{Phys. Rev. Lett.}}}
  \textbf{\bibinfo{volume}{102}}, \bibinfo{pages}{186602}
  (\bibinfo{year}{2009}).

\bibitem{Wiesendanger_NatRevMats_2016}
\bibinfo{author}{Wiesendanger, R.}
\newblock \bibinfo{journal}{\bibinfo{title}{Nanoscale magnetic skyrmions in
  metallic films and multilayers: a new twist for spintronics}}.
\newblock {\emph{\JournalTitle{Nature Reviews Materials}}}
  \textbf{\bibinfo{volume}{1}}, \bibinfo{pages}{16044} (\bibinfo{year}{2016}).

\bibitem{Fert_NatRevMats_2017}
\bibinfo{author}{Fert, A.}, \bibinfo{author}{Reyren, N.} \&
  \bibinfo{author}{Cros, V.}
\newblock \bibinfo{journal}{\bibinfo{title}{Magnetic skyrmions: advances in
  physics and potential applications}}.
\newblock {\emph{\JournalTitle{Nature Reviews Materials}}}
  \textbf{\bibinfo{volume}{2}}, \bibinfo{pages}{17031} (\bibinfo{year}{2017}).

\bibitem{Hagemeister_NatCommun_2015}
\bibinfo{author}{Hagemeister, J.}, \bibinfo{author}{Romming, N.},
  \bibinfo{author}{von Bergmann, K.}, \bibinfo{author}{Vedmedenko, E.~Y.} \&
  \bibinfo{author}{Wiesendanger, R.}
\newblock \bibinfo{journal}{\bibinfo{title}{Stability of single skyrmionic
  bits}}.
\newblock {\emph{\JournalTitle{Nature Communications}}}
  \textbf{\bibinfo{volume}{6}}, \bibinfo{pages}{8455} (\bibinfo{year}{2015}).

\bibitem{Jonietz_Science_2010}
\bibinfo{author}{Jonietz, F.} \emph{et~al.}
\newblock \bibinfo{journal}{\bibinfo{title}{Spin transfer torques in {M}n{S}i
  at ultralow current densities}}.
\newblock {\emph{\JournalTitle{Science}}} \textbf{\bibinfo{volume}{330}},
  \bibinfo{pages}{1648} (\bibinfo{year}{2010}).

\bibitem{Yu_NatCommun_2012}
\bibinfo{author}{Yu, X.~Z.} \emph{et~al.}
\newblock \bibinfo{journal}{\bibinfo{title}{Skyrmion flow near room temperature
  in an ultralow current density}}.
\newblock {\emph{\JournalTitle{Nature Communications}}}
  \textbf{\bibinfo{volume}{3}}, \bibinfo{pages}{988} (\bibinfo{year}{2012}).

\bibitem{Fert_NatNanotechnol_2013}
\bibinfo{author}{Fert, A.}, \bibinfo{author}{Cros, V.} \&
  \bibinfo{author}{Sampaio, J.}
\newblock \bibinfo{journal}{\bibinfo{title}{Skyrmions on the track}}.
\newblock {\emph{\JournalTitle{Nature Nanotechnology}}}
  \textbf{\bibinfo{volume}{8}}, \bibinfo{pages}{152} (\bibinfo{year}{2013}).

\bibitem{Iwasaki_NatNanotechnol_2013}
\bibinfo{author}{Iwasaki, J.}, \bibinfo{author}{Mochizuki, M.} \&
  \bibinfo{author}{Nagaosa, N.}
\newblock \bibinfo{journal}{\bibinfo{title}{Current-induced skyrmion dynamics
  in constricted geometries}}.
\newblock {\emph{\JournalTitle{Nature Nanotechnology}}}
  \textbf{\bibinfo{volume}{8}}, \bibinfo{pages}{742} (\bibinfo{year}{2013}).

\bibitem{Sampaio_NatNanotechnol_2013}
\bibinfo{author}{Sampaio, J.}, \bibinfo{author}{Cros, V.},
  \bibinfo{author}{Rohart, S.}, \bibinfo{author}{Thiaville, A.} \&
  \bibinfo{author}{Fert, A.}
\newblock \bibinfo{journal}{\bibinfo{title}{Nucleation, stability and
  current-induced motion of isolated magnetic skyrmions in nanostructures}}.
\newblock {\emph{\JournalTitle{Nature Nanotechnology}}}
  \textbf{\bibinfo{volume}{8}}, \bibinfo{pages}{839} (\bibinfo{year}{2013}).

\bibitem{Jiang_Science_2015}
\bibinfo{author}{Jiang, W.} \emph{et~al.}
\newblock \bibinfo{journal}{\bibinfo{title}{Blowing magnetic skyrmion
  bubbles}}.
\newblock {\emph{\JournalTitle{Science}}} \textbf{\bibinfo{volume}{349}},
  \bibinfo{pages}{283} (\bibinfo{year}{2015}).

\bibitem{Moreau-Luchaire_NatNanotechnol_2016}
\bibinfo{author}{Moreau-Luchaire, C.} \emph{et~al.}
\newblock \bibinfo{journal}{\bibinfo{title}{Additive interfacial chiral
  interaction in multilayers for stabilization of small individual skyrmions at
  room temperature}}.
\newblock {\emph{\JournalTitle{Nature Nanotechnology}}}
  \textbf{\bibinfo{volume}{11}}, \bibinfo{pages}{444} (\bibinfo{year}{2016}).

\bibitem{Boulle_NatNanotechnol_2016}
\bibinfo{author}{Boulle, O.} \emph{et~al.}
\newblock \bibinfo{journal}{\bibinfo{title}{Room-temperature chiral magnetic
  skyrmions in ultrathin magnetic nanostructures}}.
\newblock {\emph{\JournalTitle{Nature Nanotechnology}}}
  \textbf{\bibinfo{volume}{11}}, \bibinfo{pages}{449} (\bibinfo{year}{2016}).

\bibitem{Woo_NatMater_2016}
\bibinfo{author}{Woo, S.} \emph{et~al.}
\newblock \bibinfo{journal}{\bibinfo{title}{Observation of room-temperature
  magnetic skyrmions and their current-driven dynamics in ultrathin metallic
  ferromagnets}}.
\newblock {\emph{\JournalTitle{Nature Materials}}}
  \textbf{\bibinfo{volume}{15}}, \bibinfo{pages}{501} (\bibinfo{year}{2016}).

\bibitem{Legrand_NanoLett_2017}
\bibinfo{author}{Legrand, W.} \emph{et~al.}
\newblock \bibinfo{journal}{\bibinfo{title}{Room-temperature current-induced
  generation and motion of sub-100 nm skyrmions}}.
\newblock {\emph{\JournalTitle{Nano Letters}}} \textbf{\bibinfo{volume}{17}},
  \bibinfo{pages}{2703} (\bibinfo{year}{2017}).

\bibitem{Litzius_NatPhysics_2017}
\bibinfo{author}{Litzius, K.} \emph{et~al.}
\newblock \bibinfo{journal}{\bibinfo{title}{Skyrmion hall effect revealed by
  direct time-resolved x-ray microscopy}}.
\newblock {\emph{\JournalTitle{Nature Physics}}} \textbf{\bibinfo{volume}{13}},
  \bibinfo{pages}{170} (\bibinfo{year}{2016}).

\bibitem{Woo_NatCommun_2017}
\bibinfo{author}{Woo, S.} \emph{et~al.}
\newblock \bibinfo{journal}{\bibinfo{title}{Spin-orbit torque-driven skyrmion
  dynamics revealed by time-resolved x-ray microscopy}}.
\newblock {\emph{\JournalTitle{Nature Communications}}}
  \textbf{\bibinfo{volume}{8}}, \bibinfo{pages}{15573} (\bibinfo{year}{2017}).

\bibitem{Zhang_NatCommun_2018}
\bibinfo{author}{Zhang, S.~L.} \emph{et~al.}
\newblock \bibinfo{journal}{\bibinfo{title}{Manipulation of skyrmion motion by
  magnetic field gradients}}.
\newblock {\emph{\JournalTitle{Nature Communications}}}
  \textbf{\bibinfo{volume}{9}}, \bibinfo{pages}{2115} (\bibinfo{year}{2018}).

\bibitem{Komineas_PRB_2015}
\bibinfo{author}{Komineas, S.} \& \bibinfo{author}{Papanicolaou, N.}
\newblock \bibinfo{journal}{\bibinfo{title}{Skyrmion dynamics in chiral
  ferromagnets}}.
\newblock {\emph{\JournalTitle{Physical Review B}}}
  \textbf{\bibinfo{volume}{92}}, \bibinfo{pages}{064412}
  (\bibinfo{year}{2015}).

\bibitem{Wang_NJP_2017}
\bibinfo{author}{Wang, C.}, \bibinfo{author}{Xiao, D.}, \bibinfo{author}{Chen,
  X.}, \bibinfo{author}{Zhou, Y.} \& \bibinfo{author}{Liu, Y.}
\newblock \bibinfo{journal}{\bibinfo{title}{Manipulating and trapping skyrmions
  by magnetic field gradients}}.
\newblock {\emph{\JournalTitle{New Journal of Physics}}}
  \textbf{\bibinfo{volume}{19}}, \bibinfo{pages}{083008}
  (\bibinfo{year}{2017}).

\bibitem{Liang_NJP_2018}
\bibinfo{author}{Liang, J.~J.} \emph{et~al.}
\newblock \bibinfo{journal}{\bibinfo{title}{Magnetic field gradient driven
  dynamics of isolated skyrmions and antiskyrmions in frustrated magnets}}.
\newblock {\emph{\JournalTitle{New Journal of Physics}}}
  \textbf{\bibinfo{volume}{20}}, \bibinfo{pages}{053037}
  (\bibinfo{year}{2018}).

\bibitem{Burrowes_APL_2013}
\bibinfo{author}{Burrowes, C.} \emph{et~al.}
\newblock \bibinfo{journal}{\bibinfo{title}{Low depinning fields in
  {T}a-{C}o{F}e{B}-{M}g{O} ultrathin films with perpendicular magnetic
  anisotropy}}.
\newblock {\emph{\JournalTitle{Applied Physics Letters}}}
  \textbf{\bibinfo{volume}{103}}, \bibinfo{pages}{182401}
  (\bibinfo{year}{2013}).

\bibitem{Buttner_NatNanotechnol_2017}
\bibinfo{author}{B{\"u}ttner, F.} \emph{et~al.}
\newblock \bibinfo{journal}{\bibinfo{title}{Field-free deterministic ultrafast
  creation of magnetic skyrmions by spin-orbit torques}}.
\newblock {\emph{\JournalTitle{Nature Nanotechnology}}}
  \textbf{\bibinfo{volume}{12}}, \bibinfo{pages}{1040} (\bibinfo{year}{2017}).

\bibitem{Jaiswal_APL_2017}
\bibinfo{author}{Jaiswal, S.} \emph{et~al.}
\newblock \bibinfo{journal}{\bibinfo{title}{Investigation of the
  {D}zyaloshinskii-{M}oriya interaction and room temperature skyrmions in
  {W}/{C}o{F}e{B}/{M}g{O} thin films and microwires}}.
\newblock {\emph{\JournalTitle{Appl. Phys. Lett.}}}
  \textbf{\bibinfo{volume}{111}}, \bibinfo{pages}{022409}
  (\bibinfo{year}{2017}).

\bibitem{Lemesh_AdvMater_2018}
\bibinfo{author}{Lemesh, I.} \emph{et~al.}
\newblock \bibinfo{journal}{\bibinfo{title}{Current-induced skyrmion generation
  through morphological thermal transitions in chiral ferromagnetic
  heterostructures}}.
\newblock {\emph{\JournalTitle{Advanced Materials}}}
  \textbf{\bibinfo{volume}{30}}, \bibinfo{pages}{1805461}
  (\bibinfo{year}{2018}).

\bibitem{Heide_PRB_2008}
\bibinfo{author}{Heide, M.}, \bibinfo{author}{Bihlmayer, G.} \&
  \bibinfo{author}{Bl\"ugel, S.}
\newblock \bibinfo{journal}{\bibinfo{title}{Dzyaloshinskii-{M}oriya interaction
  accounting for the orientation of magnetic domains in ultrathin films:
  {F}e/{W}(110)}}.
\newblock {\emph{\JournalTitle{Phys. Rev. B}}} \textbf{\bibinfo{volume}{78}},
  \bibinfo{pages}{140403} (\bibinfo{year}{2008}).

\bibitem{Thiaville_EPL_2012}
\bibinfo{author}{Thiaville, A.}, \bibinfo{author}{Rohart, S.},
  \bibinfo{author}{Ju{\'e}, {\'E}.}, \bibinfo{author}{Cros, V.} \&
  \bibinfo{author}{Fert, A.}
\newblock \bibinfo{journal}{\bibinfo{title}{Dynamics of dzyaloshinskii domain
  walls in ultrathin magnetic films}}.
\newblock {\emph{\JournalTitle{Europhysics Letters}}}
  \textbf{\bibinfo{volume}{100}}, \bibinfo{pages}{57002}
  (\bibinfo{year}{2012}).

\bibitem{Lemesh_PRB_2017}
\bibinfo{author}{Lemesh, I.}, \bibinfo{author}{B\"uttner, F.} \&
  \bibinfo{author}{Beach, G. S.~D.}
\newblock \bibinfo{journal}{\bibinfo{title}{Accurate model of the stripe domain
  phase of perpendicularly magnetized multilayers}}.
\newblock {\emph{\JournalTitle{Phys. Rev. B}}} \textbf{\bibinfo{volume}{95}},
  \bibinfo{pages}{174423} (\bibinfo{year}{2017}).

\bibitem{Yamanouchi_IEEEMagnLett_2011}
\bibinfo{author}{Yamanouchi, M.} \emph{et~al.}
\newblock \bibinfo{journal}{\bibinfo{title}{Domain structure in cofeb thin
  films with perpendicular magnetic anisotropy}}.
\newblock {\emph{\JournalTitle{IEEE Magnetics Letters}}}
  \textbf{\bibinfo{volume}{2}}, \bibinfo{pages}{3000304}
  (\bibinfo{year}{2011}).

\bibitem{Grady_2006}
\bibinfo{author}{Grady, L.}
\newblock \bibinfo{journal}{\bibinfo{title}{Random walks for image
  segmentation}}.
\newblock {\emph{\JournalTitle{IEEE Transactions on Pattern Analysis and
  Machine Intelligence}}} \textbf{\bibinfo{volume}{28}}, \bibinfo{pages}{1768}
  (\bibinfo{year}{2006}).

\bibitem{Malek_CzechJPhys_1958}
\bibinfo{author}{M\'alek, Z.} \& \bibinfo{author}{Kambersk\'y, V.}
\newblock \bibinfo{journal}{\bibinfo{title}{On the theory of the domain
  structure of thin films of magnetically uni-axial materials}}.
\newblock {\emph{\JournalTitle{Czech. J. Phys.}}} \textbf{\bibinfo{volume}{8}},
  \bibinfo{pages}{416} (\bibinfo{year}{1958}).

\bibitem{Torrejon_NatCommun_2014}
\bibinfo{author}{Torrejon, J.} \emph{et~al.}
\newblock \bibinfo{journal}{\bibinfo{title}{Interface control of the magnetic
  chirality in {C}o{F}e{B}/{M}g{O} heterostructures with heavy-metal
  underlayers}}.
\newblock {\emph{\JournalTitle{Nature Communications}}}
  \textbf{\bibinfo{volume}{5}}, \bibinfo{pages}{4655} (\bibinfo{year}{2014}).

\bibitem{LoConte_PRB_2015}
\bibinfo{author}{Lo~Conte, R.} \emph{et~al.}
\newblock \bibinfo{journal}{\bibinfo{title}{Role of {B} diffusion in the
  interfacial {D}zyaloshinskii-{M}oriya interaction in
  {T}a/{C}o$_{20}${F}e$_{60}${B}$_{20}$/{M}g{O} nanowires}}.
\newblock {\emph{\JournalTitle{Phys. Rev. B}}} \textbf{\bibinfo{volume}{91}},
  \bibinfo{pages}{014433} (\bibinfo{year}{2015}).

\bibitem{Khan_APL_2016}
\bibinfo{author}{Khan, R.~A.} \emph{et~al.}
\newblock \bibinfo{journal}{\bibinfo{title}{Effect of annealing on the
  interfacial {D}zyaloshinskii-{M}oriya interaction in {T}a/{C}o{F}e{B}/{M}g{O}
  trilayers}}.
\newblock {\emph{\JournalTitle{Applied Physics Letters}}}
  \textbf{\bibinfo{volume}{109}}, \bibinfo{pages}{132404}
  (\bibinfo{year}{2016}).

\bibitem{Soucaille_PRB_2016}
\bibinfo{author}{Soucaille, R.} \emph{et~al.}
\newblock \bibinfo{journal}{\bibinfo{title}{Probing the
  {D}zyaloshinskii-{M}oriya interaction in {C}o{F}e{B} ultrathin films using
  domain wall creep and {B}rillouin light spectroscopy}}.
\newblock {\emph{\JournalTitle{Phys. Rev. B}}} \textbf{\bibinfo{volume}{94}},
  \bibinfo{pages}{104431} (\bibinfo{year}{2016}).

\bibitem{Ma_PRL_2018}
\bibinfo{author}{Ma, X.} \emph{et~al.}
\newblock \bibinfo{journal}{\bibinfo{title}{Interfacial
  {D}zyaloshinskii-{M}oriya interaction: {E}ffect of $5d$ band filling and
  correlation with spin mixing conductance}}.
\newblock {\emph{\JournalTitle{Phys. Rev. Lett.}}}
  \textbf{\bibinfo{volume}{120}}, \bibinfo{pages}{157204}
  (\bibinfo{year}{2018}).

\bibitem{Di_APL_2015}
\bibinfo{author}{Di, K.} \emph{et~al.}
\newblock \bibinfo{journal}{\bibinfo{title}{Asymmetric spin-wave dispersion due
  to {D}zyaloshinskii-{M}oriya interaction in an ultrathin {P}t/{C}o{F}e{B}
  film}}.
\newblock {\emph{\JournalTitle{Applied Physics Letters}}}
  \textbf{\bibinfo{volume}{106}}, \bibinfo{pages}{052403}
  (\bibinfo{year}{2015}).

\bibitem{Chen_APL_2018}
\bibinfo{author}{Chen, Y.} \emph{et~al.}
\newblock \bibinfo{journal}{\bibinfo{title}{Tuning {S}lonczewski-like torque
  and {D}zyaloshinskii-{M}oriya interaction by inserting a {P}t spacer layer in
  {T}a/{C}o{F}e{B}/{M}g{O} structures}}.
\newblock {\emph{\JournalTitle{Applied Physics Letters}}}
  \textbf{\bibinfo{volume}{112}}, \bibinfo{pages}{232402}
  (\bibinfo{year}{2018}).

\bibitem{Rohart_PRB_2013}
\bibinfo{author}{Rohart, S.} \& \bibinfo{author}{Thiaville, A.}
\newblock \bibinfo{journal}{\bibinfo{title}{Skyrmion confinement in ultrathin
  film nanostructures in the presence of {D}zyaloshinskii-{M}oriya
  interaction}}.
\newblock {\emph{\JournalTitle{Phys. Rev. B}}} \textbf{\bibinfo{volume}{88}},
  \bibinfo{pages}{184422} (\bibinfo{year}{2013}).

\bibitem{Soumyanarayanan_NatMater_2017}
\bibinfo{author}{Soumyanarayanan, A.} \emph{et~al.}
\newblock \bibinfo{journal}{\bibinfo{title}{Tunable room-temperature magnetic
  skyrmions in {I}r/{F}e/{C}o/{P}t multilayers}}.
\newblock {\emph{\JournalTitle{Nature Materials}}}
  \textbf{\bibinfo{volume}{16}}, \bibinfo{pages}{898} (\bibinfo{year}{2017}).

\bibitem{Nagaosa_NatNanotechnol_2013}
\bibinfo{author}{Nagaosa, N.} \& \bibinfo{author}{Tokura, Y.}
\newblock \bibinfo{journal}{\bibinfo{title}{Topological properties and dynamics
  of magnetic skyrmions}}.
\newblock {\emph{\JournalTitle{Nature Nanotechnology}}}
  \textbf{\bibinfo{volume}{8}} (\bibinfo{year}{2013}).

\bibitem{Zhang_APL_2018}
\bibinfo{author}{Zhang, S.} \emph{et~al.}
\newblock \bibinfo{journal}{\bibinfo{title}{Direct writing of room temperature
  and zero field skyrmion lattices by a scanning local magnetic field}}.
\newblock {\emph{\JournalTitle{Applied Physics Letters}}}
  \textbf{\bibinfo{volume}{112}}, \bibinfo{pages}{132405}
  (\bibinfo{year}{2018}).

\bibitem{Bacani_2018}
\bibinfo{author}{Ba\'cani, M.}, \bibinfo{author}{Marioni, M.~A.},
  \bibinfo{author}{Schwenk, J.} \& \bibinfo{author}{Hug, H.~J.}
\newblock \bibinfo{journal}{\bibinfo{title}{How to measure the local
  {D}zyaloshinskii-{M}oriya interaction in skyrmion thin-film multilayers}}.
\newblock {\emph{\JournalTitle{arXiv:1609.01615}}}  (\bibinfo{year}{2018}).

\bibitem{Maccariello_NatNanotechnol_2018}
\bibinfo{author}{Maccariello, D.} \emph{et~al.}
\newblock \bibinfo{journal}{\bibinfo{title}{Electrical detection of single
  magnetic skyrmions in metallic multilayers at room temperature}}.
\newblock {\emph{\JournalTitle{Nature Nanotechnology}}}
  \textbf{\bibinfo{volume}{13}}, \bibinfo{pages}{233--237}
  (\bibinfo{year}{2018}).

\end{thebibliography}
%\noindent LaTeX formats citations and references automatically using the bibliography records in your .bib file, which you can edit via the project menu. Use the cite command for an inline citation, e.g.  \cite{Hao:gidmaps:2014}.
%For data citations of datasets uploaded to e.g. \emph{figshare}, please use the \verb|howpublished| option in the bib entry to specify the platform and the link, as in the \verb|Hao:gidmaps:2014| example in the sample bibliography file.

\section*{Acknowledgements}
%Acknowledgements should be brief, and should not include thanks to anonymous referees and editors, or effusive comments. Grant or contribution numbers may be acknowledged.
The authors thank the EMRP Joint Research Projects 15SIB06 NanoMag and 17FUN08 TOPS for financial support. In particular, A.C. thanks the Researcher Mobility Grant 15SIB06-RMG4. The EMRP is jointly funded by the EMRP participating countries within EURAMET and the European Union. The work in Mainz was supported by the Deutsche Forschungsgemeinschaft (DFG, German Research Foundation) - project number 290319996/TRR173. We are grateful to Craig Barton and Marco Co\"isson for useful discussions.

\section*{Author contributions}
%Must include all authors, identified by initials, for example:
%A.A. conceived the experiment(s),  A.A. and B.A. conducted the experiment(s), C.A. and D.A. analysed the results.  All authors reviewed the manuscript. 
A.C. conceived the experiment. M.V. grew the samples. A.C. and M.V. characterised the samples. A.C. and H.C.L. conducted the experiments. A.C., H.C.L., and F.G.S. analysed the results. A.C. wrote the manuscript. All authors discussed the results and reviewed the manuscript.

\section*{Additional information}
%To include, in this order: \textbf{Accession codes} (where applicable); 
\textbf{Supplementary information} accompanies this paper at

\noindent \textbf{Competing interests:} The authors declare no competing interests. 
%The corresponding author is responsible for submitting a \href{http://www.nature.com/srep/policies/index.html#competing}{competing interests statement} on behalf of all authors of the paper. This statement must be included in the submitted article file.

\end{document}